\documentstyle[12pt]{article}
\begin{document}
\vspace{3cm}
\rightline{\bf\large {JHU-TIPAC-930016}}
\vspace{1cm}
{\centerline {\bf \LARGE Multi Particle Semiclassical Process}}
\vspace{0.5cm}
{\centerline {\bf \LARGE  in $\phi^{4}$ Theory\footnote{This work supported
 in part by the National Science Foundation under grant PHY-90-96198.}}}
\vspace{1cm}
{\centerline {\bf \large Alexander
Kyatkin\footnote{kyatkin@dirac.pha.jhu.edu}}}
\vspace{1cm}
{\centerline {\em Department of Physics and Astronomy}}
{\centerline {\em The Johns Hopkins University}}
{\centerline {\em Baltimore, MD 21218, USA}}
\vspace{1cm}
{\centerline {\bf\large {May 1993}}}
\vspace{1cm}
\begin{abstract}
We have shown an example of semiclassical transition in
$\phi^{4}$ model with positive coupling constant. This process describes
 a semiclassical transition between two coherent states with much
  smaller average number of particles in the initial state than in the final
 state. This transition
 is technically analogous to the one-instanton transition in the
 electroweak model. It is suppressed by the factor $\exp(-2S_{0})$
, where $S_{0}$ is  Lipatov instanton action.  It could be important
 to the problem of calculation
 of amplitudes for multiparticle production in $\phi^4$-type models.
\end{abstract}
\newpage
\noindent {\bf 1.}\ \ \ Recently,
considerable efforts have been made to calculate
amplitudes for multiparticle production in  weakly coupled
field theories. The study of this problem
 was initiated by the observation of the fact \cite {Ringwald}
 that baryon-number violating processes in the electroweak
 theory, associated with multiparticle production, could become
 relevant at energy scale $E\sim 10\,TeV$. This problem gave
 impulse to
 study multiparticle amplitudes in the simpler case of $\phi^4$
 model \cite {Cornwall,Shifman}
, considered before in the context of large orders
 of perturbation theory \cite {Lipatov}.

The semiclassical methods for computing such amplitudes in
 the electroweak theory use Euclidean classical solutions
of the equations of motion --  instantons \cite {tHooft}.
 To apply the similar calculations to $\phi^4$ theory \cite {Shifman},
 we have to use Lipatov's trick \cite {Lipatov} and consider first
 the theory with negative coupling constant. Then, the theory with
 negative coupling constant allows a
 semiclassical instanton-like transitions.

In this paper we show that $\phi^4$ theory allows
 a semiclassical transition even for the case of
positive coupling constant. This transition is described by
 a classical $O(4)$-invariant solution, considered on a contour
 in the complex time plane. The ``type'' of the transition is
 determined by the position of this contour with respect to the positions
 of the singularities of the classical solution.

 The transition is technically
 analogous to the one-instanton
 transitions in the electroweak model. It
is suppressed by the factor $\exp {(-2S_{0})}$, where $S_{0}$
is equal to Lipatov  instanton action -- the action of
the classical solution in the Euclidean theory with  negative
coupling constant \cite {Lipatov}.

This process describes a classically-forbidden
 transition between two coherent states with a much smaller number
of particles in the initial state than in the final state
 -- $n_{final}\sim n^{5/7}_{initial}/\lambda^{2/7}$
(where $\lambda$ is a small coupling constant). Therefore, it
could be relevant to the calculation of amplitudes for multiparticle
 production in $\phi^4$-type models.
We suppose that the contribution of such a process must be included
 into the corresponding multiparticle amplitude and, probably, can
 slow down the factorial growth of the
perturbative amplitude \cite {Cornwall}.

\vspace{1cm}

\noindent {\bf 2.}\ \ \  Formally, we cannot calculate
 the transition probability for the process $two\rightarrow many$
particles in the
semiclassical manner at all, because of the
 non-semiclassical nature of the
initial two-particle state. Instead, as proposed in Ref.\cite
{Approach}, we can calculate the probability of
transition between a semiclassical initial
 state with a ``small'' number of
particles and a final semiclassical state with a ``large''
 number of particles.
The probability of such a transition can be considered
as some approximation to the two particle cross section
 in one-instanton sector and gives us an
 upper bound for this cross section.

 The starting point of this approach is the amplitude for
 a transition at fixed energy $E$ from
the initial coherent state $\mid\lbrace a_{\bf k} \rbrace
\rangle $ (projected onto this energy)
 to the final coherent state $\mid \lbrace b_
{\bf k} \rbrace \rangle $ \cite {Rubakov3}
\begin{equation}
\label{a}
A=\langle\lbrace b_{\bf k}\rbrace
\mid SP_{E}\mid \lbrace a_
{\bf k} \rbrace \rangle ,
\end{equation}
where the operator $P_{E}$
 is a projector onto subspace of
definite energy $E$; $S$ is the $S$-matrix.

When $E \sim 1/\lambda$
and $a_{\bf k}, b_{\bf k} \sim 1/\sqrt{\lambda}$ for small coupling
constant $\lambda$,
we can evaluate  the transition amplitude (1) in the saddle-point
 approximation.

In the saddle-point approximation the functional integral (1) is
dominated by the solution of the classical field equations with
some specific boundary conditions \cite {Rubakov3},
determined by the initial coherent
 state at early time $t\rightarrow -\infty$ and the
final coherent states at late time $t\rightarrow +\infty$.

\begin{picture}(300,200)
\put (50,100){\vector(1,0){200}} \put (150,40){\vector(0,1){130}}
\thicklines \put (70,140){\line(1,0){80}}
\thicklines \put (150,140){\line(0,-1){40}}
\thicklines \put (150,100){\line(1,0){80}}
  \put (185,105){$C$}
\put (157,135){$T$} \put (155,165){Im~$t$}
\put (240,87){Re~$t$} \put (80,150){A} \put (140,0){Fig.1}
\end{picture}

However, for the calculation of some classically-forbidden transition
 (for example tunneling) we cannot restrict ourselves to a pure
Minkowski or Euclidean time, because we make deal simultaneously
 with the classically-allowed event (such as
free evolution of the initial and final states)
 and classically-forbidden event. So we work with the  contour
in the complex time plane shown on Fig. (1) \cite {Rubakov3}.

The part A of this contour is shifted upward
 and runs parallel to the real axis $t=t^{\prime}+iT$.
 Evolution of the system with respect to $t^{\prime}$
corresponds to initial state propagation,  while the real
 part of the contour describes final state propagation.

The boundary value problem is conveniently formulated on this contour
\cite {Rubakov3}.
We assume below that the classical solution
$\phi$ becomes free at large initial
 and final time, which means that its spatial Fourier transform
 can be written as a superposition of plane waves. Then, the negative
frequency part of the classical
field , considered on the part A of the contour
 at the limit $t^\prime\rightarrow -\infty$,
is determined by the initial state. The positive frequency part
of the field is determined by the final state on Minkowski part
of the contour at large positive time $t\rightarrow +\infty$.

Thus, to find the transition probability,
we have to solve the field equations
 with fixed negative frequency part of the field
at early time and positive frequency part of the field
at late time. This is an extremely difficult problem for
 arbitrary initial and final states, even in the
case of the $\phi^4$ theory. So we are forced to restrict
ourselves to a less general problem, proposed in Ref. \cite
 {Rubakov4}: we find first {\em some} real Minkowski-time
 solution and then find the corresponding initial and final
states as asymptotics of this solution.

We have to make some remarks about the choice of the ``appropriate''
 solution.

First, we consider only real solutions
 because, as it has been shown in Ref.\cite {Rubakov4},
  the probability of the transition from the given initial state
 to all possible final states is saturated by a single final state
 which is real at real time. Therefore, the real
 saddle-point configuration
 corresponds to the transition from the given initial state to the most
probable final state.

The second condition is that this solution should
have an appropriate singularity structure
 in the complex time plane - we have to be able to choose the contour
 of Fig.(1) and avoid any singularities of the solution.

 We will show below that $\phi^4$ theory possesses
 such solutions.

The semiclassical suppression in the transition probability
 is determined by the imaginary part of the classical action,
 calculated along the time contour of Fig. (1) \cite {Rubakov4}
 (see also \cite {me})
\begin{equation}
\label{b}
\sigma=\mid A\mid^{2}\,\sim\,\exp(-2{\rm Im}\,S).
\end{equation}

The probability of the transition does not depend on the
choice of the contour and we can move the contour upward or downward
 until we reach a
 singularity of the classical solution. So the ``type'' of the transition
 is determined by the position of the contour with respect
 to the position of the singularities of
 the classical solution in the complex time plane.

\vspace{1cm}

\noindent {\bf 3.}\ \ \  Now we apply this formalism to $\phi^4$ model.

The action of the model (we consider a
real scalar field), written
in conformally invariant form \cite {conform}, is
$$
S=\int d^4 x\, (-{1\over 2}\phi\,\partial_{\mu}\partial^{\mu}\phi-
{\lambda\over 4}\,{\phi}^{4}) ,
$$
where $\lambda>0$ is the small coupling constant. The corresponding
classical field equation is
\begin{equation}
\label{m}
\partial^2 \phi+\lambda\,\phi^3=0
\end{equation}

$O(4)$-invariant solutions of this equation \cite {Castell} can
be easily found using the invariance of the massless theory under the
Minkowski conformal group. This invariance
 can be made explicit by projecting
the theory onto the surface of the hypertorus \cite {Schechter}. Then,
$O(4)$-invariant solutions can be found by solving a one-dimensional
equation and they correspond to the oscillations with
 amplitude $a$ in
the one-dimensional potential $V(x)=x^2/2+\lambda
x^4/4$.

 The $O(4)$-invariant solution can be expressed in terms of elliptic
functions
\begin{equation}
\label{n}
\phi(\vec x,t)={1\over \sqrt{\lambda}}
{{2a}\over {\sqrt {(r^2-(t-i)^2)(r^2-(t+i)^2)}}}\,
{\rm cn}(\sqrt {1+a^2}\,\zeta-\zeta_{0},{\it k}^2) ,
\end{equation}
where $r=\mid\vec {x}\mid$, ${\it k}^2={a^2/(2(1+a^2))}$
and
$$
\zeta={1\over {2i}}\, \ln({{r^2-(t-i)^2}\over {r^2-(t+i)^2}}) .
$$
Here {\bf cn} stands for the Jacobi elliptic cosine (see, for example,
 \cite {elliptic}) and {\it k} is
the modulus of this function. The arbitrary integration constants are
$a$ and $\zeta_{0}$. We choose $\zeta_{0}=K$ (where $K=\int_{0}^{\pi\over 2}
{{dx}/{\sqrt{
1-{\it k}^2\, \sin^{2}x}}}$ is the complete elliptic integral), in which
case $\phi=0$ at $t=0$. The constant $a$, as we will see below,
 is related to the
energy.

According to the above described approach,
 we are going to calculate
a transition corresponding to the saddle point configuration (4),
 considered
on the time contour of Fig.(1) for some value of parameter $T={\rm Im}\,t$.

First, we investigate
 the analytic structure of the solution in the complex time plane.

This solution is real on the real time axis , so, as has been
shown in Ref. \cite {Rubakov4}, it corresponds to a transition
from the given initial state to the most probable final state.

The solution has essential ``light-cone''
singularities at $t={\pm} x {\pm} i$.
Hence,  we have to choose $T<1$ for the contour of Fig.(1) not to cross
the ``light-cone'' singularity.

In addition, there are
singularities (poles) at the ``points'' where
$$
\sqrt {1+a^2}\,\zeta -K=2mK+(2n+1)i\,K^{\prime}.
 $$
Here $K^{\prime} (k^2)=K(1-k^2)$, $m,n=0,\pm 1,\pm 2,...$.
 These ``points'' are poles
 of the elliptic cosine \cite {elliptic}. Because $\zeta$ is
 a function of radial coordinate and complex time, the solutions
 of this equation determine the singularity curves in the coordinate
 axes $r$, ${\rm Re}\,t$, ${\rm Im}\,t$.

We will consider below only the case $a<<1$, which, as will be shown
below, corresponds to the case of a ``small'' number of
final-state
particles ($n_{final}<<{1/\lambda}$). In this limit
 $K\approx\,{\pi/2}$
 and only $m=-1$ and $n\geq{0}$ case
corresponds to the singularities in
the region ${\rm Im}\,t\geq 0$, ${\rm Re}\,t\leq 0$.
 The singularities curves (numerated by
integer number $n$) $t=t_{n}(r)$
run asymptotically ``parallel'' to the ``light-cone''
 and have $({\rm Im}\,t)$ coordinate close
 to 1
\begin{equation}
\label{o}
t_{n}=i\,\bigl(1-{({a^2\over 16})}^{2n+1}\bigr)-r
\end{equation}
at $r\rightarrow+\infty$ and $n=0,1,2...$. We have shown in Fig. (2)
  two curves in the region
 ${\rm Im}\,t\geq 0$, ${\rm Re}\,t\leq 0$. The structure of singularities
 of the classical solution is similar to the structure of singularities
 of particular classical solutions in
 two-dimensional $\sigma$-model and four
dimensional Yang-Mills theory, investigated
 in this context in Ref.\cite {Rubakov4} and \cite {TE}.

We can see that the structure of the singularities
of this solution is ``appropriate'' --
we are able to choose the contour of Fig.(1) and not to cross
 any singularities. We choose the contour with
exactly one singularity curve under it (i.e. with $1-{a^2/16}<
T<1-({a^2/16})^3$). It will be shown below that this choice
corresponds to a classically-forbidden (exponentially suppressed)
transition.

\begin{picture}(300,300)
\put (200,100){\vector(1,0){100}}
\put (200,100){\vector(0,1){195}}
\put (200,100){\vector(-1,-1){64}}
\put (150,180){\line(-3,-1){90}}
\put (150,210){\line(-3,-1){90}}
\put (150,190){\oval(20,20)[br]}
\put (150,220){\oval(20,20)[br]}
\put (153,211){\line(-3,-1){10}}
\put (153,181){\line(-3,-1){10}}
\put (200,275){\line(-3,-1){140}}
\put (200,275){\line(-1,0){30}}
\put (150,275){\line(-1,0){30}}
\put (100,275){\line(-1,0){30}}
\put (160,190){\line(-1,0){20}}
\put (130,190){\line(-1,0){20}}
\put (100,190){\line(-1,0){20}}
\put (160,220){\line(-1,0){20}}
\put (130,220){\line(-1,0){20}}
\put (100,220){\line(-1,0){20}}
\put (90,213){\circle*{2}}
\put (90,223){\circle*{2}}
\put (90,233){\circle*{2}}
\put (70,197){$t_{1}(r)$}
\put (70,167){$t_{0}(r)$}
\put (145,35){$r$}
\put (290,87){Re~$t$}
\put (205,285){Im~$t$}
\put (193,20){Fig.~2}
\put (70,260){"$light-cone$"}
\end{picture}

The leading suppression factor in the
transition probability (2) is determined by the imaginary part of the
 action calculated along the contour of Fig.(1).
 To calculate the imaginary part of the action we use the method of
Ref.\cite {Rubakov4}.

The action of the model is
$$
S={\lambda\over 2}\int\,d^3x\,\int\limits_{C}\,dt\,\phi^4(\vec {x},t) ,
$$
where we have used the equation of motion.  For every ${\bf x}$ the time
 integral along the contour of Fig. (1) is equal to
 to the sum of the
integral along the real time axis (which is real) and contribution of
the pole $t_{0}$, corresponding to the
singularity (5) at $n=0$. The pole contribution can be calculated
using the expression for the {\bf cn} near the singularity $-2K+iK^{\prime}$
 \cite {elliptic}
$$
{\rm cn}\,(-2K+i\,K^{\prime}+u)=-{1\over iku}-{1\over 6ik}\,(1-2k^2)\,u+O(u^2)
$$
and expanding $\zeta$ in Taylor series up to the fourth order.

After lengthy calculations we obtain for the imaginary part of the action
$$
{\rm Im}S={8\pi^2\over 3\lambda}.
$$
It is exactly equal to Lipatov instanton action: the
 Euclidean action of the classical solution in $\phi^4$ theory with
 negative coupling constant \cite {Lipatov} (our normalization of
$\lambda$ differs from the normalization of $\lambda$ in \cite {Lipatov}
by factor 6).
 Thus, the choice of the contour between the first and the
second singularity line corresponds to the classically
forbidden transition suppressed by the factor
$$
\sigma\,\sim\,\exp(-2\,S_{0}) ,
$$
where $S_{0}$ is equal to Lipatov instanton action.
 So this process is analogous to the one-instanton transition in
the electroweak model or to the ``instanton-like'' transition in $\phi^4$
 theory with negative coupling constant \cite {Lipatov}.

Our calculation of the transition probability has some resemblance
 to Landau approach to the calculation of the classically-forbidden
reflection from a potential barrier in one-dimensional quantum
mechanics \cite {Landau} (for energy larger then the height of
 barrier). In Landau approach the reflection probability is
 also determined by the imaginary
part of the classical action on a trajectory
in the complex coordinate plane, ``wound'' around a singular ``turning''
 point.

So the evolution along the imaginary part of the contour of Fig. (1)
can be interpreted
 as a classically-forbidden reflection in the $\phi^4$ potential.
Part A of the contour describes the free propagation
 of an incoming spherically-symmetric shell (4) at early time.
The
 Minkowski part of the contour corresponds to an outgoing wave
 at late time. Therefore,
  we can call the
semiclassical process in the $\phi^4$ model, described by the
nontrivial ``trajectory'' (lying between the singular lines)
in the complex time plane, as a
``transmission after classically-forbidden reflection''.

\vspace{1cm}

\noindent {\bf 4.}\ \ \ As has been mentioned
before, the final state is determined
 via Eq.(6) by the asymptotics of the classical solution on the Minkowski
part of the contour of Fig.(1) in the limit $t\rightarrow +\infty$, while
 the initial state corresponds to the asymptotics of the solution on
 part A of the contour in the limit $t^\prime\rightarrow -\infty$,
 where $t=iT+t^\prime$. The average numbers of particles
  are determined by the negative frequency Fourier components $g_{\bf k}$
  of the classical
 solution (4), considered on the corresponding parts of the contour
$$
n=
\int\,d{\bf k}\,g_{\bf k}^{\ast}\,g_{\bf
  k}.
$$

We analyze only case $a<<1$ which, we will see below, corresponds
to the case $n_{final}<<{1/\lambda}$.

In this limit we obtain for the average number of the final
 particle
$$
n_{final}={\pi^2\,a^2\over \lambda}  .
$$

The number of the initial particles is given by
$$
n_{initial}=
({3\over 4^4})^{1/5}{\pi^2\,
a^{14/5}\over 2\,\lambda}
={1\over 2}\,({3\over 4^4})^{1/5}\,a^{4/5}\,n_{final} .
$$
This result implies
$$
n_{final}\sim\,{n_{initial}^{5/7}\over \lambda^{2/7}}.
$$

We can also define the average momentum \cite {Rubakov4}
$$
{\rm k}_{average}
\approx\,{E/n}
$$
(energy in terms of amplitude $a$ is expressed
 by $E={\pi^2\,a^2\over \lambda}$). Then, the initial average momentum
 is related with the final momentum by
$$
{\rm k}_{average}^{initial}\sim\,{{\rm k}_{average}^{final}\over a^{4/5}}.
$$

We can see that for small coupling constant the number
of the final ``soft'' particles is much larger then the number of
the initial ``hard'' particles.

Thus, the classical solution, considered on the contour of Fig.(1) in
the complex time plane above the singularity line, corresponds to the
 transition between two coherent states with a ``strong''
 violation of particle number, $n_{final}>>n_{initial}$.

\vspace{1cm}

\noindent {\bf 5.}\ \ \ To conclude, we have studied the
semiclassical process in $\phi^4$ theory with positive
coupling constant, which describes transition
between two coherent states. This transition is suppressed
 by the factor
$\exp (-2S_{0})$, where $S_{0}$ is equal to the Lipatov
instanton action -- the Euclidean action of the classical solution
 in the  theory with negative coupling constant.

The initial and final states, corresponding to this transition
, have different numbers of particles ($n_{final}>>n_{initial}$)
 and different average momenta (${\rm k}_{final}<< {\rm k}_{initial}$),
 so this transition approximates some multiparticle
scattering process with a large number
 of ``soft'' final particles.

 The process is technically analogous to the one-instanton
 transition
in electroweak model and could serve as a
good model for studying the instanton effects. It seems that we can
also describe some ``multi-instanton'' processes using the solution
 (4) and choosing the contour of Fig.(1) above several singularity
lines.

We believe that we have to include the contributions
 of these instanton-like
processes into the corresponding ``total'' amplitude for multiparticle
production. Such contributions might slow down the
 factorial growth of the perturbative amplitude
and unitarize the high energy cross section.

The energy dependence of the transition probability
 of this process is a very interesting problem.
The growth of the transition probability
is related to the presence of the external
particles.
 An accurate estimation of the contribution
 of the external particles requires, however,
  including mass term effects
 into consideration.

We have to add the mass term for the following
  reason.
  Calculation of the
 transition probability requires summing the
contributions from different ``sizes'' of the
classical field. In this paper we consider only the contribution
 of the solution with
 a ``unit'' size (field configuration (4)).
 This integration is divergent at the large
 ``sizes'' and should be regularized by introducing
  a mass term into the action in the manner of the ``constrained
 instanton'' approach \cite {Affleck}.
We do not consider the effects of the mass term in this paper, so
 this problem requires a more detailed investigation.

The important point is that the
 framework of the formalism allows one to analyze, in principle,
  the most interesting
case $n_{final}\ge {1/\lambda}$. This case is analogous
 to multi-particle scattering at the sphaleron energy
in the standard model, where the behavior of multi-particle
 cross section is still far from being understood.

\vspace{1cm}

 The author is indebted to
 T. Gould and E. Poppitz for stimulating
discussions and critical remarks, especially
regarding the problem of calculation of the
Fourier components of the initial state.
 I am grateful to J. Bagger and G. Domokos for
valuable discussions and interesting suggestions.

\end{document}